\documentstyle[emulateapj]{article}

\lefthead{MENOU ET AL.} 
\righthead{RADIO BAL QUASARS IN THE SDSS}

\def\gsim{\;\rlap{\lower 2.5pt
 \hbox{$\sim$}}\raise 1.5pt\hbox{$>$}\;}
\def\lsim{\;\rlap{\lower 2.5pt
   \hbox{$\sim$}}\raise 1.5pt\hbox{$<$}\;} 

\begin{document}

\title{Broad Absorption Line Quasars In The Sloan Digital Sky Survey
With VLA-FIRST Radio Detections}

\author{Kristen Menou\altaffilmark{\ref{Princeton}}, Daniel E. Vanden
Berk\altaffilmark{\ref{Fermilab}}, \v{Z}eljko
Ivezi\'{c}\altaffilmark{\ref{Princeton}}, Rita
S.J. Kim,\altaffilmark{\ref{Princeton}} Gillian R.
Knapp\altaffilmark{\ref{Princeton}}, Gordon T.
Richards\altaffilmark{\ref{PSU}}, Iskra
Strateva\altaffilmark{\ref{Princeton}}, Xiaohui
Fan\altaffilmark{\ref{IAS}}, James
E. Gunn\altaffilmark{\ref{Princeton}}, Patrick
B. Hall\altaffilmark{\ref{Princeton},\ref{Catol}}, Tim
Heckman\altaffilmark{\ref{JHU}}, Julian
Krolik\altaffilmark{\ref{JHU}}, Robert H.
Lupton\altaffilmark{\ref{Princeton}}, Donald
P. Schneider\altaffilmark{\ref{PSU}}, Donald
G. York\altaffilmark{\ref{Chicago},\ref{Chicago2}},
S.F. Anderson\altaffilmark{\ref{Washington}},
N.A. Bahcall\altaffilmark{\ref{Princeton}},
J. Brinkmann\altaffilmark{\ref{APO}},
R. Brunner\altaffilmark{\ref{Caltech}},
I. Csabai\altaffilmark{\ref{JHU}},
M. Fukugita\altaffilmark{\ref{CosmicRay}},
G.S. Hennessy\altaffilmark{\ref{Naval}},
P.Z. Kunszt\altaffilmark{\ref{JHU}},
D.Q. Lamb\altaffilmark{\ref{Chicago}},
J.A. Munn\altaffilmark{\ref{Flagstaff}},
R.C. Nichol\altaffilmark{\ref{CMU}},
G.P. Szokoly\altaffilmark{\ref{Potsdam}} }

\newcounter{address}
\setcounter{address}{1}
\altaffiltext{\theaddress}{Princeton University Observatory, 
Princeton, NJ 08544
\label{Princeton}}
\addtocounter{address}{1}
\altaffiltext{\theaddress}{Fermi National Accelerator Laboratory, P.O. Box 500,
Batavia, IL 60510
\label{Fermilab}}
\addtocounter{address}{1}
\altaffiltext{\theaddress}{Department of Astronomy \& Astrophysics,
Pennsylvania State University,
University Park, PA 16802
\label{PSU}}
\addtocounter{address}{1}
\altaffiltext{\theaddress}{Institute for Advanced Study, Olden Lane,
Princeton, NJ 08540
\label{IAS}}
\addtocounter{address}{1}
\altaffiltext{\theaddress}{
Pontificia Universidad Cat\'{o}lica de Chile, Departamento de Astronom\'{\i}a
y Astrof\'{\i}sica, Facultad de F\'{\i}sica, Casilla 306, Santiago 22, Chile
\label{Catol}}
\addtocounter{address}{1}
\altaffiltext{\theaddress}{
Department of Physics \& Astronomy, The Johns Hopkins University,
   3701 San Martin Drive, Baltimore, MD 21218, USA
\label{JHU}}
\addtocounter{address}{1}
\altaffiltext{\theaddress}{The University of Chicago, Astronomy \& Astrophysics
Center, 5640 S. Ellis Ave., Chicago, IL 60637
\label{Chicago}}
\addtocounter{address}{1}
\altaffiltext{\theaddress}{Enrico Fermi Institute, 5640 So. Ellis Ave., Chicago, IL 60615
\label{Chicago2}}
\addtocounter{address}{1}
\altaffiltext{\theaddress}{University of Washington, Department of Astronomy,
Box 351580, Seattle, WA 98195
\label{Washington}}
\addtocounter{address}{1}
\altaffiltext{\theaddress}{Apache Point Observatory, P.O. Box 59,
Sunspot, NM 88349-0059
\label{APO}}
\addtocounter{address}{1}
\altaffiltext{\theaddress}{Department of Astronomy, California Institute
of Technology, Pasadena, CA 91125
\label{Caltech}}
\addtocounter{address}{1}
\altaffiltext{\theaddress}{Institute for Cosmic Ray Research, University of
Tokyo, Midori, Tanashi, Tokyo 188-8502, Japan
\label{CosmicRay}}
\addtocounter{address}{1}
\altaffiltext{\theaddress}{U.S. Naval Observatory, 3450 Massachusetts 
Avenue NW, Washington, DC 20392-5420
\label{Naval}}
\addtocounter{address}{1}
\altaffiltext{\theaddress}{U.S. Naval Observatory, Flagstaff Station, 
P.O. Box 1149, 
Flagstaff, AZ  86002-1149
\label{Flagstaff}}
\addtocounter{address}{1}
\altaffiltext{\theaddress}{Department of Physics, Carnegie Mellon University,
     5000 Forbes Ave., Pittsburgh, PA-15232
\label{CMU}}
\addtocounter{address}{1}
\altaffiltext{\theaddress}{Astrophysikalisches Institut Potsdam, An
der Sternwarte 16, D-14482 Potsdam, Germany
\label{Potsdam}}

\begin{abstract}
We present 13 Broad Absorption Line (BAL) quasars, including 12 new
objects, which were identified in the Sloan Digital Sky Survey (SDSS)
and matched within $2''$ to sources in the FIRST radio survey
catalog. The surface density of this sample of radio-detected BAL
quasars is $4.5 \pm 1.2$ per $100~$deg$^2$, i.e. approximately 4 times
larger than previously found by the shallower FIRST Bright Quasar
Survey (FBQS).  A majority of these radio-detected BAL quasars are
moderately radio-loud objects.  The fraction of BAL quasars in the
entire radio quasar sample, $4.8\pm 1.3\%$, is comparable to the
fraction of BAL quasars among the SDSS optical quasar sample (ignoring
selection effects). We estimate that the true fraction of BAL quasars
(mostly HiBALs) in the radio sample is $9.2\pm 2.6\%$ once selection
effects are accounted for.  We caution that the absorption troughs of
4~of the 13~radio-detected quasars considered do not strictly satisfy
the standard BALnicity criterion.  One or possibly two of the new
radio-detected BAL quasars are of the rare ``iron LoBAL'' type. BAL
quasars are generally redder than the median SDSS quasar at the same
redshift.

\end{abstract}


\section{Introduction}

The spectra of Broad Absorption Line (BAL) quasars show broad,
blueshifted absorption troughs corresponding to highly-ionized
restframe-UV transitions such as C~{\sc iv}, Si~{\sc iv}, N~{\sc v},
O~{\sc vi}, more rarely Mg~{\sc ii} and Al~{\sc iii} and even more
rarely Fe~{\sc ii} (e.g.  Hazard et al. 1987; Weymann et al. 1991;
Becker et al. 1997). The similarity of the continuum and line emission
of BAL and non-BAL quasars motivates the hypothesis that BAL quasars
are not intrinsically different from other quasars. The presence of
BAL features in the spectra of only $\sim 10\%$ of optically-selected
quasars could naturally be explained by a difference in viewing angle
if the subrelativistic outflow at the origin of the BAL features is
not isotropic (Weymann et al. 1991). The popular notion that the
outflow is preferentially located in the plane of the disk surrounding
the supermassive black hole has found support in spectro-polarimetric
measurements for BAL quasars (e.g. Goodrich \& Miller 1995; Cohen et
al. 1995; see, e.g., Murray et al. 1995 for a theoretical wind model).

This geometrical interpretation of BAL features suggests the possible
existence of links between the BAL characteristics of a quasar and its
radio properties (such as jet orientation, radio spectral index or
even radio-loudness).  Along those lines, Stocke et al. (1992) found
that their sample of optically-selected BAL quasars revealed only
radio-quiet sources, while BAL quasars were strikingly absent from a
sample of radio-loud objects. The authors proposed that the BAL
phenomenon is simply anticorrelated with the mechanism giving rise to
strong radio emission in quasars; they also noted the tendency for
radio-loud quasars at the high-end of the radio-loudness distribution
to lack high-velocity BALs (see also Weymann et al. 1991). Other
interpretations of the BAL phenomenon have been presented, such as the
evolutionary scenario of Briggs, Turnshek \& Wolfe (1984; see also
Gregg et al. 2000) in which BAL quasars later become radio-loud
objects.

The mJy flux-limited VLA-FIRST survey (Becker, White \& Helfand 1995)
has challenged many of the standard views concerning radio quasars and
the BAL phenomenon. In the FIRST Bright Quasar Survey (FBQS; Gregg et
al. 1996; White et al. 2000), FIRST sources with point-source optical
counterparts brighter than 17.8 in the APM catalog POSS-I plates were
systematically targeted for spectroscopic identification. The FBQS,
besides challenging the existence of a clear radio-dichotomy for
quasars (White et al. 2000), has established beyond doubt the
existence of a population of radio-loud BAL quasars (Becker et
al. 2000; see also Hazard et al. 1987; Becker et al. 1997, Brotherton
et al. 1998; Wills, Brandt \& Laor 1999). Furthermore, the population
of radio-detected BAL quasars identified by the FBQS was found to
exhibit a diversity of radio spectral indices (both steep and flat
spectra) which may not support an interpretation of BAL quasars as
preferentially edge-on oriented systems (Becker et al. 2000; see also
Gregg et al. 2000).

In this paper, we describe preliminary results for BAL quasars
spectroscopically identified by the Sloan Digital Sky Survey (SDSS)
and matched within $2''$ to radio counterparts in the FIRST survey
catalog. These results are part of a more ambitious project aimed at
studying all SDSS sources with FIRST radio detections (see Ivezi\'{c}
et al. 2001; Knapp et al. 2001).  Our main result is that, at the
greater depth probed by SDSS in the optical, the surface density of
radio-detected BAL quasars is $\sim 4$ times larger than previously
reported by the FBQS, with a majority of moderately radio-loud
objects.

In \S2, we describe how SDSS and FIRST sources were selected to be
part of the sample of BAL quasars discussed. The sample and some of
its characteristics are described in \S3, while results concerning the
optical (SDSS photometric) properties are emphasized in \S4. We
conclude in \S5.

\section{Target Selection}

\subsection{SDSS Photometric and Spectroscopic Systems}

The Sloan Digital Sky Survey (SDSS; York et al. 2000) uses a camera
with an array of CCDs (Gunn et al. 1998) on a dedicated 2.5-m
telescope (Siegmund et al. 2001) at Apache Point Observatory, New
Mexico, to obtain images in five broad optical bands over
10,000~deg$^2$ of the high Galactic latitude sky ($b \gsim 30^{\rm
o}$) centered approximately on the North Galactic Pole.  The five
filters (designated $u'$, $g'$, $r'$, $i'$, and $z'$) cover the entire
wavelength range of the CCD response, longward of the atmospheric
cutoff at short wavelengths (Fukugita et al. 1996).  Photometric
calibration is provided by simultaneous observations with a 20-inch
telescope at the same site.  The survey data processing software
measures the properties of each detected object in the imaging data,
and determines and applies astrometric and photometric calibrations
(Pier et al. 2001; Lupton et al. 2001).\footnote{SDSS nomenclature:
the names for sources have the format \hbox{SDSSp
Jhhmmss.ss+ddmmss.s}, where the coordinate equinox is~J2000, and the
``p" refers to preliminary.  The reported magnitudes are asinh
magnitudes (Lupton, Gunn \& Szalay 1999) and are based on a
preliminary photometric calibration; to indicate this the filters have
an asterisk instead of a prime superscript.  The estimated current
astrometric accuracies in each coordinate are~0.1$''$ and the
calibration of the photometric measurements are currently accurate
to~0.02--0.05~magnitudes.}

Based on their optical morphology and colors, a subset of the sources
discovered in the imaging survey is selected for spectroscopic
follow-up.  The spectroscopic survey uses two fiber--fed double
spectrographs designed to cover the wavelength range 3800--9200
{\AA}. The SDSS spectrographs achieve a spectral resolution of $\sim$
1800 across the entire spectral range. Each spectrograph accepts 320
fibers, each of which subtends a diameter of 3$^{''}$ on the
sky. Exposures are typically $\sim 45-60$ minutes long. Details
concerning the SDSS spectroscopic system are given by Uomoto et
al. (2001) and Frieman et al. (2001).

Quasar candidates are selected from their point-source optical
morphology\footnote{At low redshift, this morphology cut is not
enforced so that extended AGN are included in the spectroscopic sample
as well (Richards et al. 2001b).} and optical colors such that they
lie outside the stellar locus in color-color space (blue colors for $z
\lsim 2$ quasars -- see box in Fig.~\ref{fig:two}a -- and red colors
for $z \gsim 3$ quasars; Fan 1999; Fan et al. 1999). In addition,
point sources with radio counterparts within $2''$ in the FIRST
catalog are favored candidates for spectroscopy (because there is very
little stellar contamination in radio matches, independent of the
optical colors; Helfand et al. 1999; Ivezi\'{c} et al. 2001; Knapp et
al. 2001).  The quasar selection criteria used for this sample was not
constant throughout the period of observations; however, all otherwise
good quasar candidates with reddening corrected $i^*$ magnitudes
brighter than $19$ were flagged as spectroscopic targets.  In
addition, other quasar candidates (mostly high-redshift candidates)
were selected as faint as $i^* = 20.5$.  Radio-detected quasar
candidates were selected to the brighter of these limits, unless they
were also high-redshift candidates (see Richards et al. 2001b for a
detailed account on the selection for spectroscopic follow-up).

\subsection{Matching to the VLA-FIRST Survey Catalog}

The identification in the FIRST catalog of several hundreds of radio
counterparts to SDSS optical sources shows that there is very little
contamination by spurious associations at angular separations $\lsim
2''$ (Ivezi\'{c} et al. 2001). The FIRST catalog provides both peak
and integrated 1.4~GHz flux densities down to 1~mJy (corresponding to
a $ \sim 5 \sigma$ detection). The source radio morphology can be
studied from the FIRST survey images.\footnote{FIRST website: {\tt
http://sundog.stsci.edu}} For comparison, the flux limit for the radio
sample of Stocke et al. (1992) is typically 0.2-0.3~mJy.

\section{Results}

\subsection{The Sample of Quasars}

We report on results concerning 60~SDSS spectroscopic plates, covering
a total sky area of $\approx 290$~deg$^2$. The total area was
calculated by summing the areas corresponding to individual
spectroscopic plates and by taking into account the overlaps between
the plates.  The completeness of the radio sample is expected to be
nearly $100\%$ because only spectroscopic plates corresponding to
regions covered by the VLA-FIRST survey were considered.  Overall, a
total of 2326~targets were spectroscopically identified as quasars
(see Vanden Berk et al. 2001 for definition; see also Richards et
al. 2001a), 96 of which show BAL features as determined by visual
inspection. A subset of 272 quasars have reliable FIRST detections
(i.e. radio counterparts within $2''$), while 13 of the 96 BAL quasars
do.\footnote{In what follows, we refer to objects without a FIRST
radio detection as ``radio-undetected'' and objects with a FIRST radio
detection as ``radio-detected''.} The focus of this paper is on these
13~radio-detected BAL objects, whose properties are listed in
Table~\ref{tab:one}: J2000 coordinates, SDSS optical magnitudes $u^*$,
$g^*$, $r^*$, $i^*$, $z^*$ (reddening-corrected), peak radio flux in
the FIRST catalog, BALnicity index (see definition below), maximum
outflow velocity $V_{\rm max}$, a representative absolute magnitude,
$M_{g^*}$, radio-loudness $R^*$ (k-corrected ratio of radio to optical
fluxes), 1.4~GHz specific radio luminosity $L_{\rm R}$ (in
erg$^{-1}$~s$^{-1}$~Hz$^{-1}$), redshift $z$ and BAL classification.

The absolute magnitudes and specific radio luminosities were
calculated assuming the following cosmology: $\Omega_m=1$,
$\Omega_\lambda=0$ ($q_0=0.5$) and $H_0=50$~km~s$^{-1}$~Mpc$^{-1}$.
The "BALnicity" index is a measure of the strength of an absorption
trough, similar to an equivalent width, but requiring continuous
absorption of at least 10\% in depth and spanning at least
$2000$~km~s$^{-1}$, and excluding the region within $3000$~km~s$^{-1}$
of the quasar emission redshift (Weymann et al. 1991). C~{\sc iv} is
used in all cases for which it is accessible and Mg~{\sc ii} is used
otherwise.  The quantity "$V_{\rm max}$" is the maximum outflow
velocity of the absorption lines, relative to the quasar emission
redshift. We only list peak flux densities because of their close
agreement with the integrated values for all the sources, as expected
for radio point sources. The compact radio morphology of the 13~BAL
quasars listed was confirmed by inspecting the FIRST survey images for
each object.

A NASA/IPAC Extragalactic Database (NED) search reveals that one of
the objects (SDSSp J115944.81+011207.1) can be identified with PKS
J1159+0112, that another one (SDSSp J003923.20-001452.7) was detected
as a radio source in the NVSS survey (with a compact radio morphology
and a peak flux density consistent with the FIRST values) and that
none of the objects listed has a counterpart in the IRAS survey
catalog. No counterparts were found in the ROSAT All Sky Survey or
2MASS catalogs either (only 8 of the 13 listed objects are located in
the sky region covered by the 2MASS Second Incremental Data
Release\footnote{Available at {\tt
http://www.ipac.caltech.edu/2mass/releases/second/index.html}}).

{Following Becker et al. (2000), we do not limit our BAL sample to
the conservative definition of Weymann et al. (1991), which explains
why four of the objects listed have a zero BALnicity index (i.e. if an
absorption feature satisfied all the criteria of Weymann et al. except
being beyond $3000$~km~s$^{-1}$ of the quasar emission redshift, the
object was still considered as a potential BAL quasar). The
justifications given by Becker et al. (2000) for including objects
which do not strictly satisfy the BALnicity criterion are as follows:
(i) nearly black absorption spanning $\sim$ 4000~km~s$^{-1}$, which is
unlikely to break up into narrow lines, (ii) several absorption
systems with a large $V_{\rm max}$ value; none in excess of
2000~km~s$^{-1}$, but taken together they are suggestive of an
intrinsic BAL outflow, (iii) evidence for partial covering, which is a
property of BAL quasars and (iv) low-ionization lines in BAL quasars
tend to be narrower than high-ionization lines (Voit, Weymann \&
Korista 1993), so that any BALnicity index which had to be calculated
from low-ionization lines (because high-ionization lines could not be
observed) are likely to be a lower limit. These same criteria motivate
us to include four objects with a zero BALnicity index in our sample
of radio-detected BAL quasars. SDSSp J003923.20-001452.7 has a
measured BALnicity index $> 0$, but rounded down to 0; it satisfies
criterion (i) and maybe criterion (iii). SDSSp J235702.55-004824.0
satisfies criterion (i) and maybe criterion (ii). SDSSp
J115944.81+011207.1 satisfies criterion (i) except for $V_{\rm max}
\sim 3000$~km~s$^{-1}$; it satisfies criterion (ii): there is an
additional broad absorption system at $V_{\rm max} \sim
8000$~km~s$^{-1}$; it also appears to satisfy criterion (iii): the
C~${\sc iv}$ absorption trough is nearly flat-bottomed with a residual
flux indicating partial coverage. SDSSp J133903.41-004241.2 is the
least convincing of its category; Mg~${\sc ii}$ absorption is black
with $V_{\rm max} = 1650$~km~s$^{-1}$ , but the noisy C~${\sc iv}$
part of the spectrum may allow it to satisfy criterion (iv).  }

The radio-loudness $R^*$ is calculated as the k-corrected ratio of the
5~GHz radio flux to the 2500~$\AA$ optical flux (which is calculated
from the B-band magnitude), following the definition of Stocke et
al. (1992). We adopt the photometric transformation $B \simeq
g^*+0.13$ which is appropriate for the power-law spectra of quasars
(see, e.g., Schmidt, Schneider \& Gunn 1995) and we assume a power-law
index $\alpha_r=-0.5$ (with $f_{\nu} \propto \nu^{\alpha}$) in the
radio (the same value of the radio power-law index was assumed for the
k-correction when calculating $L_{\rm R}$). A power-law index
$\alpha_o=-0.5$ was assumed when calculating the absolute magnitudes
(see Vanden Berk et al. 2001 for a measure of $\alpha_o$ from an SDSS
quasar composite spectrum; see also Richards et al. 2001a).

The redshifts were determined automatically by the SDSS spectroscopic
pipeline and modified as necessary upon visual inspection (BAL quasar
redshifts prove difficult to measure automatically with precision but
the modifications were generally not more than a few hundredths of a
unit redshift). The BAL classification refers to the ionization states
seen in absorption: HiBALs are defined as those exhibiting absorption
only by high-ionization ions, LoBALs show both high- and
low-ionization lines, and FeLoBALs are LoBALs with lines from
meta-stable Fe~{\sc ii} and Fe~{\sc iii}. Examples of specific lines
are Mg~{\sc ii}$~\lambda 2800$ for LoBAL (as well as, e.g., lines of
Al~{\sc ii}, Al~{\sc iii}, Si~{\sc ii}, and C~{\sc ii}), C~{\sc
iv}$~\lambda 1549$ for HiBAL, and Fe~{\sc ii}$~\lambda
1063,~2261,~2380,~2600,~2750$ and Fe~{\sc iii}$~\lambda 1122,~1914$
for FeLoBAL (Hazard et al. 1987). Of course, all the lines cannot be
seen for all the quasars because of the finite spectral coverage in
the observer frame.

We note that one or possibly two of the 13~radio-detected BAL quasars
reported here are of the rare FeLoBAL type, of which the FBQS has
found only four examples (Becker et al. 2000) and only one other has
been reported (Hazard et al. 1987).  Five of the six known FeLoBALs
have radio detections, which supports the speculation by Becker et
al. (1997) that these quasars are associated with rather strong radio
emission. We note, however, that the number of FeLoBALs in the
radio-undetected sample is presently unknown and that these results
still concern a small number of sources. The spectra of the
13~radio-detected BAL quasars listed in Table~\ref{tab:one} are shown
in Figure~\ref{fig:one}.

\subsection{General Properties}

Despite the small size of this sample of radio-detected BAL quasars,
some interesting conclusions can be drawn. The surface density of
radio-detected BAL quasars at the depth probed by SDSS is $4.5 \pm
1.2$ objects per 100~deg$^2$, which is approximately $4$ times larger
than reported by the FBQS (with the same definition for the BAL
sample; Becker et al. 2000). This is expected given that the SDSS goes
deeper than the limiting magnitude of 17.9 for the FBQS.  Adopting a
dividing line between radio-quiet and radio-loud objects at $R^*=10$
(following Stocke et al. 1992), we see that most of the objects
identified by SDSS and found in the FIRST catalog are of a moderately
radio-loud ($10 < R^* < 100$) to strongly radio-loud ($R^* > 100$)
nature.  This is the result of the greater depth probed by SDSS in the
optical (as compared to the APM digitized POSS-I catalog used for the
FBQS), which, given the mJy flux limit of the FIRST survey, tends to
uncover objects which are more radio-loud than previously known. The
specific radio luminosity $L_{\rm R}$ alone is also sometimes used as
a direct measure of the radio-loudness of quasars. Adopting a dividing
line between radio-quiet and radio-loud objects at $L_{\rm R} =
10^{32.5} h_{50}^{-2}$~erg$^{-1}$~s$^{-1}$~Hz$^{-1}$ (following Gregg
et al. 1996 and Stern et al. 2000), we see again that a majority of
the objects listed in Table~\ref{tab:one} are on the radio-loud
side.\footnote{Note that a FIRST source with a 1~mJy flux (the survey
limit) must be at a redshift larger than $3$ or so to automatically
satisfy the above luminosity-based radio-loudness criterion. Although
two of our objects have $z>3$, both of them have flux densities four
times brighter than the FIRST survey limit.} A more conservative cut
at $10^{33}$~erg$^{-1}$~s$^{-1}$~Hz$^{-1}$ still results in 7 out of
the 13~objects being on the radio-loud side.  Thus, we find no
evidence of the anticorrelation of the radio-loud emission mechanism
with the BAL phenomenon which was found by Stocke et al. (1992). The
trend of moderate radio-loudness in our sample of radio-detected BAL
quasars may, however, be consistent with the tendency found by Stocke
et al. (1992; see also Weymann et al. 1991) for strongly radio-loud
quasars to lack high-velocity BALs.

One must temper the conclusion of this analysis, however, by a few
words of caution.  In particular, we note that the BALnicity
distribution of the radio-detected BAL quasars listed in
Table~\ref{tab:one} appears different (at a $\approx 2 \sigma$ level)
from that of the sample of radio-quiet BAL quasars of Weymann et al.
(1991). While $\sim 25 \%$ of our BAL quasars have a BALnicity index
$> 3000$~km~s$^{-1}$, $\sim 70 \%$ of the BAL quasars discussed by
Weymann et al.  (1991) do. These authors warn that quasars with a
BALnicity index $< 1500$~km~s$^{-1}$ risk contamination by unusually
strong ``associated narrow absorbers'' (as opposed to {\it bona fide}
BAL features), which may be preferentially found in radio-louder
objects.

{While nine of our radio-detected BAL quasars have a BALnicity
index $< 1500$~km~s$^{-1}$, we note that those with a non-zero
BALnicity index have large enough $V_{\rm max}$ values to avoid
contamination by associated narrow absorbers. The motivation for
including the remaining four objects with zero BALnicity indices was
given in \S3.1. The overall low BALnicity of the sample is
evident\footnote{We note that the absorption troughs in SDSSp
J115944.81+011207.1 and SDSSp J235702.55-004824.0 could be
characterized as associated absorption lines in the sense that they
are contained within the corresponding broad emission line. On the
other hand, FIRST sources being preferentially flat-spectrum sources,
these two objects may not necessarily belong to the class of systems
with associated absorption lines and usually steep radio spectra (see,
e.g., Hamann et al. 2001).} and it is interesting that the objects
with zero BALnicity indices are among the radio-loudest ones.  For
this reason, we also quote radio-detected BAL fractions excluding the
4~systems with a zero BALnicity index and refer to them in what
follows as ``conservative values''.  Interestingly, only $\sim 30 \%$
of the radio-detected BAL quasars found by the FBQS have a BALnicity
index $> 3000$~km~s$^{-1}$, in agreement with our results. The
discrepancy with the Weymann et al. sample may therefore indicate a
real difference between the populations of radio-quiet and radio-loud
BAL quasars.  We also note that the redshift distribution of our
radio-detected BAL quasars appears different from that of the FBQS
radio BAL quasars: none of our objects is found below $z \sim 1.5$,
while 11 of the 29 FBQS objects are in this low redshift range.  }

Our results concerning various interesting fractions for the quasar
sample are summarized in Table~\ref{tab:two}.  In the total SDSS
quasar sample, $11.7 \pm 0.7\%$ of all spectroscopically identified
quasars have radio detections in the FIRST survey, in agreement with a
fraction $13.5 \pm 3.8\%$ for the BAL subsample only. The fraction of
BAL quasars in the entire SDSS quasar sample is $4.1 \pm 0.4\%$, which
is consistent with the fraction $4.8 \pm 1.3\%$ (conservative value:
$3.3 \pm 1.1\%$) of BAL quasars in the subsample of sources with radio
detection. These values suggest that BAL quasars are not
preferentially found among radio-detected quasars (in contradiction to
Becker et al. 2000 for the FBQS), but are found in equal number among
sources with and without a FIRST radio detection. This apparent
discrepancy may be explained by the necessity for Becker et al. (2000)
to compare their BAL quasar fraction ({\it after} correction for the
selection effects) to those obtained in independent optical
surveys. The combination of SDSS optical data and FIRST radio data
should provide a more robust estimate for the radio-detected and
radio-undetected BAL fractions; a larger number of sources is required
to settle this issue with confidence.

The fractions of BAL quasars quoted above do not take into account the
strong selection effects which affect the identification of BAL
quasars (the restframe-UV absorption features must be conveniently
located in the optical spectral range observed). Accounting for these
selection effects is usually done by assuming that there is no
significant redshift dependence of the BAL quasar fraction. Under this
assumption, we calculate the corrected fraction as the ratio of the
number of BAL quasars to the total number of quasars in the redshift
range such that the detection of a specific restframe-UV absorption
line is guaranteed given the SDSS spectral coverage (see, e.g., Becker
et al. 2000 for a similar procedure for the FBQS).

The redshift range relevant to LoBAL quasars is $0.4 \lsim z \lsim
1.7$ based on Mg~{\sc ii} absorption. However, we find that Al~{\sc
iii} absorption is present in nearly all the spectra showing Mg~{\sc
ii} absorption. This effectively pushes the maximum detection redshift
for LoBALs to $z \approx 3.9$. This situation illustrates how the
uniform signal to noise and extended spectral coverage of the SDSS
will allow to define more sophisticated BAL fractions (based on
several specific lines) which should be less subject to selection
effects than usual. A total of 249~radio-detected quasars, including 7
LoBAL quasars, are found in the redshift range $0.4 \lsim z \lsim 3.9$
(based on both Mg~{\sc ii} and Al~{\sc iii} absorption), so that the
corrected LoBAL fraction is $ 2.8 \pm 1.1 \%$ (conservative value:
$2.0 \pm 0.9 \%$) in the sample of radio-detected objects.

The redshift range relevant to HiBAL quasars is $z \gsim 1.4$, so that
C~{\sc iv} absorption would be seen if present. A total of
141~radio-detected quasars, including 13 HiBAL quasars (all the
quasars listed in Table~\ref{tab:one} show high-ionization BAL
features), are found in this redshift range, so that the corrected
HiBAL fraction is $9.2 \pm 2.6 \%$ (conservative value: $\sim 6.4 \pm
2.1 \%$) in the sample of radio-detected objects.  For the FBQS,
Becker et al. (2000) find a corrected LoBAL fraction of $\sim 3\%$,
which is consistent with the $ \sim 2.8\%$ above, but a corrected
HiBAL fraction of $18 \pm 3.8\%$, which is significantly higher than
the $9.2 \pm 2.6\%$ HiBAL fraction of the present radio sample.  We
note that the HiBAL fraction in optical samples such as the LBQS is
consistent with the $\sim 9\%$ quoted here (e.g. Foltz et al. 1990).
The same exercise for the radio-undetected BAL quasars requires a
careful BAL classification for these objects, which is beyond the
scope of this preliminary communication.

\subsection{Optical Properties}

The optical properties of the sample of radio-detected BAL quasars are
further described in Figures~\ref{fig:two} and~\ref{fig:three}, where
similar properties for the sample of radio-undetected BAL quasars are
also shown. Fig~\ref{fig:two}a and~\ref{fig:two}b illustrate how most
of the 96 BAL quasars were selected for spectroscopy, based on their
location in the ($u^*-g^*$, $g^*-r^*$, $r^*-i^*$) SDSS color space. In
one color plane or the other, most BAL quasars lie outside the stellar
locus, indicated by contours and small dots (for individual objects)
for an SDSS control sample of $40,000$ point sources, almost all of
which are stars. The photometric quality of the control sample was
guaranteed by applying the following cuts: $u^*,~g^*,~r^*< 21$ and
corresponding photometric errors $< 0.1$.  The BAL quasars only fill
the red side of the low-z quasar box (indicated by dashed lines in
Fig.~\ref{fig:two}a) because of their rather large redshifts ($z \gsim
1$) and comparatively red colors (see below).  The low-$z$ quasar box
approximates the region of color-space where the majority of $z<2.2$
quasars are found.  The circled objects (radio-detected BAL quasars)
are evenly distributed among the BAL quasar sample in
Fig.~\ref{fig:two}, showing that there is no detectable color
difference between the radio-detected and radio-undetected objects.

Figure~\ref{fig:three}a shows a redshift-magnitude ($i^*$) diagram for
the BAL quasar sample, showing again no obvious trend separating the
radio-detected sources from the radio-undetected ones. Since the
standard photometric cut applied in SDSS for spectroscopic follow-up
is at $i^* < 19$, Fig.~\ref{fig:three}a suggests that the sample of
radio-detected BAL quasars is not strongly biased toward faint
magnitudes (relative to the rest of the BAL sample) as could have
resulted from the favored spectroscopic selection for radio-detected
sources.

Figure~\ref{fig:three}b shows a redshift-color ($g^*-i^*$) diagram for
the BAL quasar samples. The $g^*-i^*$ color was chosen so that
systematics appearing when strong emission lines enter or leave
photometric bands (especially the $r^*$ band; Richards et al. 2001a)
are minimized. The relation for the median color as a function of
redshift found by Richards et al. (2001a) for a nearly identical
sample of 2625 spectroscopically identified quasars with SDSS colors
is shown as a solid line, while the dashed lines indicate the $95\%$
confidence limits. The location of many BAL quasars above the median
value (solid line) clearly shows that BAL quasars are redder than the
typical quasars spectroscopically identified by SDSS, in agreement
with previous results on the colors of BAL quasars (Sprayberry \&
Foltz 1992; Brotherton et al. 2000; see also Weymann et al. 1991).
This red nature is interpreted as continuum extinction rather than
attributed to the presence of the absorption troughs in the BAL quasar
spectra (Sprayberry \& Foltz 1992; Yamamoto \& Vansevicius 1999). This
was confirmed here for the 13~radio-detected BAL quasars by inspection
of the spectra and comparison to the SDSS composite spectrum of Vanden
Berk et al. (2001).  We note that the red colors of BAL quasars could
imply that they are strongly under-represented in magnitude-limited
samples. Further work is required to address this question properly.

Since $\sim 15\%$ of the radio-undetected objects are below the median
$g^*-i^*$ color, $\sim 2$ of the radio-detected BAL quasars are
expected below the median value if the two populations have similar
color properties. One cannot conclude, however, from the absence of
any radio-detected BAL quasar below the solid line that they are
redder than radio-undetected BAL quasars because the samples are too
small at this point to draw statistically significant
conclusions. Interestingly, Richards et al. (2001a) found that many of
the reddest quasars in their sample have FIRST radio detections.

\section{Conclusion}

We described the properties of 13~BAL quasars spectroscopically
identified by the SDSS which possess radio counterparts in the
VLA-FIRST survey catalog. This sample, from an area of $\approx
290$~deg$^2$, is the second largest sample of radio-detected BAL
quasars after that reported by the FIRST Bright Quasar Survey (Becker
et al. 2000).

Despite the small statistical size of this sample, we were able to
isolate some of its important characteristics. Because of the SDSS
limiting magnitude of $\sim 19-20$ for our sample, compared to $17.9$
for the FIRST Bright Quasar Survey, we find a density of
radio-detected BAL quasars of $\sim 4.5$ objects per $100$~deg$^2$,
i.e. four times larger than in the FIRST Bright Quasar Survey.  A
majority of the newly identified radio-detected BAL quasars are
moderately radio-loud objects, in contradiction with early claims of
an anti-correlation between the radio-loud emission mechanism and the
BAL phenomenon. The preference for moderate radio-loudness in our
sample may indicate that strongly radio-loud quasars tend to lack high
velocity BALs.

Upon completion, the combination of the full set of SDSS data with the
FIRST catalog should provide us with several hundreds of
radio-detected BAL quasars, allowing a robust characterization of the
properties of this population of rare objects.

\section*{Acknowledgments}

The Sloan Digital Sky Survey (SDSS) is a joint project of The
University of Chicago, Fermilab, the Institute for Advanced Study, the
Japan Participation Group, The Johns Hopkins University, the
Max-Planck-Institute for Astronomy, New Mexico State University,
Princeton University, the United States Naval Observatory, and the
University of Washington. Apache Point Observatory, site of the SDSS
telescopes, is operated by the Astrophysical Research Consortium
(ARC).  Funding for the project has been provided by the Alfred
P. Sloan Foundation, the SDSS member institutions, the National
Aeronautics and Space Administration, the National Science Foundation,
the U.S. Department of Energy, Monbusho, and the Max Planck
Society. The SDSS Web site is http://www.sdss.org/.

This research has made use of the NASA/IPAC Extragalactic Database
(NED) which is operated by the Jet Propulsion Laboratory, California
Institute of Technology, under contract with the National Aeronautics
and Space Administration.

We are grateful to M. Strauss for useful discussions, R. Becker for
comments on the manuscript and the referee for a very useful report.
Support for this work was provided by NASA through Chandra
Postdoctoral Fellowship grant number PF9-10006 awarded (to KM) by the
Chandra X-ray Center, which is operated by the Smithsonian
Astrophysical Observatory for NASA under contract NAS8-39073. GRK and
IS acknowledge support from NASA grant NAG-3364. DPS and GTR
acknowledge support from NSF grant AST99-00703.

\clearpage

\begingroup
\tabcolsep=3pt
\scriptsize
\begin{deluxetable}{rrcccccrrrrrrrc}
\tablewidth{0pt}
\tablecaption{SDSS Broad Absorption Line Quasars with FIRST Detections}
\tablenum{1}
\tablehead{
\colhead{RA (J2000)}&\colhead{Dec (J2000)}&\colhead{$u^*$}&\colhead{$g^*$}&\colhead{$r^*$}&\colhead{$i^*$}&\colhead{$z^*$}&\colhead{$S_p$}&\colhead{BALnicity\tablenotemark{a}}&\colhead{$V_{\rm max}$\tablenotemark{a}}&\colhead{$M_{r^*}$}&\colhead{$R^*$}&\colhead{$\log(L_{\rm R})$}&\colhead{z\tablenotemark{a}}&\colhead{BAL}\\
&&&&&&&(mJy)&(km~s$^{-1}$)&(km~s$^{-1}$)&&&&&Type\\
\colhead{(1)}&\colhead{(2)}&\colhead{(3)}&\colhead{(4)}&\colhead{(5)}&\colhead{(6)}&\colhead{(7)}&\colhead{(8)}&\colhead{(9)}&\colhead{(10)}&\colhead{(11)}&\colhead{(12)}&\colhead{(13)}&\colhead{(14)}&\colhead{(15)}
}
\startdata
00 39 23.20&$-$00 14 52.7& 20.85 & 20.39 & 20.07 & 19.75 &19.41 & 21.2& 0       &  5100 &$-$25.15 & 419.3 &33.62&2.233& Hi\tablenotemark{b}\nl
03 05 43.45&$-$01 06 22.1& 21.13 & 20.05 &19.89 &19.54 &19.55 & 5.2& 400 &  12150 & $-$25.99 &  69.4&33.21&2.850&Hi\nl
11 54 04.14&$+$00 14 19.5& 18.85 &18.18 &17.9 &17.76 & 17.73  & 1.5& 4100 & 16200 & $-$26.68 & 4.2 &32.18&1.604&  Lo\nl
11 59 44.81&$+$01 12 07.1& 18.30 &17.48 &17.23 &16.99 &16.72 & 266.5&   0 &  3000 &$-$27.82 & 377.4 &34.63&1.989& Hi\tablenotemark{c}\nl
\smallskip
13 02 08.27&$-$00 37 31.6& 18.75 &18.46 &17.93 &17.59 &17.57 & 11.2& 1200 & 10050 & $-$26.48 &  41.3 &33.11&1.672&  Lo\nl
13 21 39.86&$-$00 41 52.0& 23.37 &20.37 &19.25 &18.65 &18.41 & 4.1& 3500 & 12150 &$-$25.83 & 70.4&33.17&3.080&FeLo\nl
13 23 04.58&$-$00 38 56.7& 18.61 & 18.54 & 18.24 &17.81 &17.77 & 8.9& 300    &   9600 & $-$26.58 & 34.6 &33.08&1.821&Hi\nl
13 31 50.52&$+$00 45 18.8& 19.28 &19.26&19.14 &18.91 &18.9 & 2.9& 1300 & 12900 &$-$25.94 & 21.9 &32.63&1.892& Hi\nl
13 39 03.41& $-$00 42 41.2& 22.14 &21.63 &21.02 &20.27 &20.01& 2.0& 0 &   1650 &$-$23.11 & 142.6 &32.28&1.518&Lo \tablenotemark{d}\nl
\smallskip
14 01 12.01& $+$01 11 12.3& 20.28&19.89 &19.59 & 19.10& 19.02& 3.1& 1500 &  10350 &$-$25.17 & 41.6 &32.59&1.771&Lo\nl
15 16 36.78&$+$00 29 40.5& 19.83 &18.51 &17.65 &17.26 &17.12 & 2.2& 6700 & 25000 & $-$27.05 & 7.6 &32.64&2.248&FeLo?\nl
16 04 12.39&$-$00 08 07.9&  20.74 &19.61 &19.42 &19.12 &18.99 & 1.4&  600 & 7200 &$-$26.42 & 12.6 &32.64&2.832& Hi\nl
23 57 02.55& $-$00 48 24.0&  22.17 &19.38 &19.03 &18.78 &18.53 & 4.0&  0  &  4050 &$-$26.77 & 28.4 &33.15&3.005&  Lo \tablenotemark{e}\nl
\tablenotetext{a}{The BALnicity values have been rounded to the nearest $100$~km~s$^{-1}$ and the $V_{\rm max}$ values to the nearest $150$~km~s$^{-1}$ to reflect the estimated precision. The estimated uncertainties in the redshift measurements (based on MgII if available or CIII] otherwise) are $< 0.005$ for all the objects, except SDSSp~J132139.86-004152.0 for which the redshift measurement is accurate only to 0.02.}
\tablenotetext{b}{NED: NVSS source J003923-001452; possibly MgII BAL. Measured BALnicity is $>$ 0 (rounded here), but value is consistent with 0 given spectral resolution and flux density errors.}
\tablenotetext{c}{NED: identified with PKS J1159+0112; possibly MgII BAL; $V_{\rm max} < 5000$~km~s$^{-1}$.}
\tablenotetext{d}{used MgII BAL; $V_{\rm max} < 5000$~km~s$^{-1}$.}
\tablenotetext{e}{$V_{\rm max} < 5000$~km~s$^{-1}$.}

\enddata
\label{tab:one}
\end{deluxetable}
\endgroup

\clearpage

\begingroup
\tabcolsep=3pt
\footnotesize
\begin{deluxetable}{ll}
\tablewidth{0pt}
\tablecaption{Radio and BAL Fractions in the SDSS Quasar Sample}
\tablenum{2}
\tablehead{
\colhead{Sub-sample}&\colhead{Fraction}\\
}
\startdata
Quasars with Radio Detection:~ & $11.7 \pm 0.7\%$ \nl
BAL Quasars with Radio Detection:~ & $13.5 \pm 3.8\%$ \nl
Quasars with BAL Features (uncorrected):~ &  $4.1 \pm 0.4\%$ \nl
Radio Quasars with BAL Features (uncorrected):~ &  $4.8 \pm 1.3\%$ ($3.3 \pm 1.1\%$)\tablenotemark{a}\nl
Radio Quasars with LoBAL Features (corrected):~ &  $2.8 \pm 1.1 \%$ ($2.0 \pm 0.9 \%$)\tablenotemark{a}\nl
Radio Quasars with HiBAL Features (corrected):~ &  $9.2 \pm 2.6\%$ ($6.4 \pm 2.1 \%$)\tablenotemark{a}\nl
\tablenotetext{a}{The fraction in parenthesis corresponds to the value  obtained when a conservative BAL definition is adopted (see text).}
\enddata
\label{tab:two}
\end{deluxetable}
\endgroup

\setcounter{figure}{0}

\clearpage

\begin{figure}
\plotone{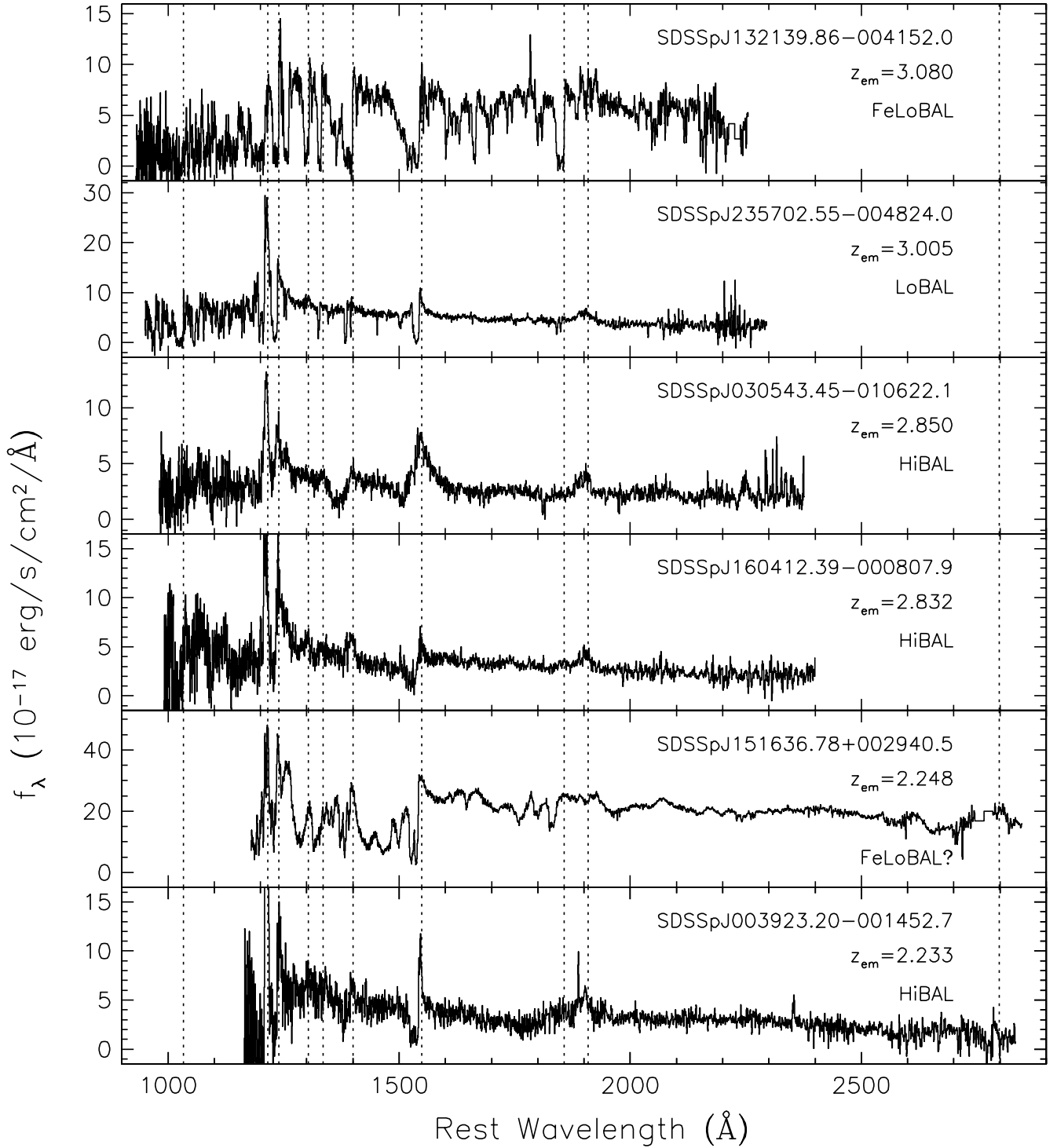}
\caption{Mosaic of rest-frame SDSS spectra for the 13 BAL quasars with
FIRST detections (SDSSp J115944.81+011207.1 can be identified with PKS
J1159+0112).  The location of Ly$\beta$/O~{\sc vi}, Ly$\alpha$, N~{\sc
v}, O~{\sc i}/Si~{\sc ii}, C~{\sc ii}, Si~{\sc iv}/O~{\sc iv}, C~{\sc
iv}, Al~{\sc iii}, C~{\sc iii}] and Mg~{\sc ii} emission lines are
indicated by dotted lines.  The spectra have a nearly constant
resolution of approximately 1800. They were all smoothed over 3
pixels, except for the lowest redshift one which was smoothed over 7
pixels.
\label{fig:one}}
\end{figure}

\clearpage
\begin{figure}
\addtocounter{figure}{-1}
\plotone{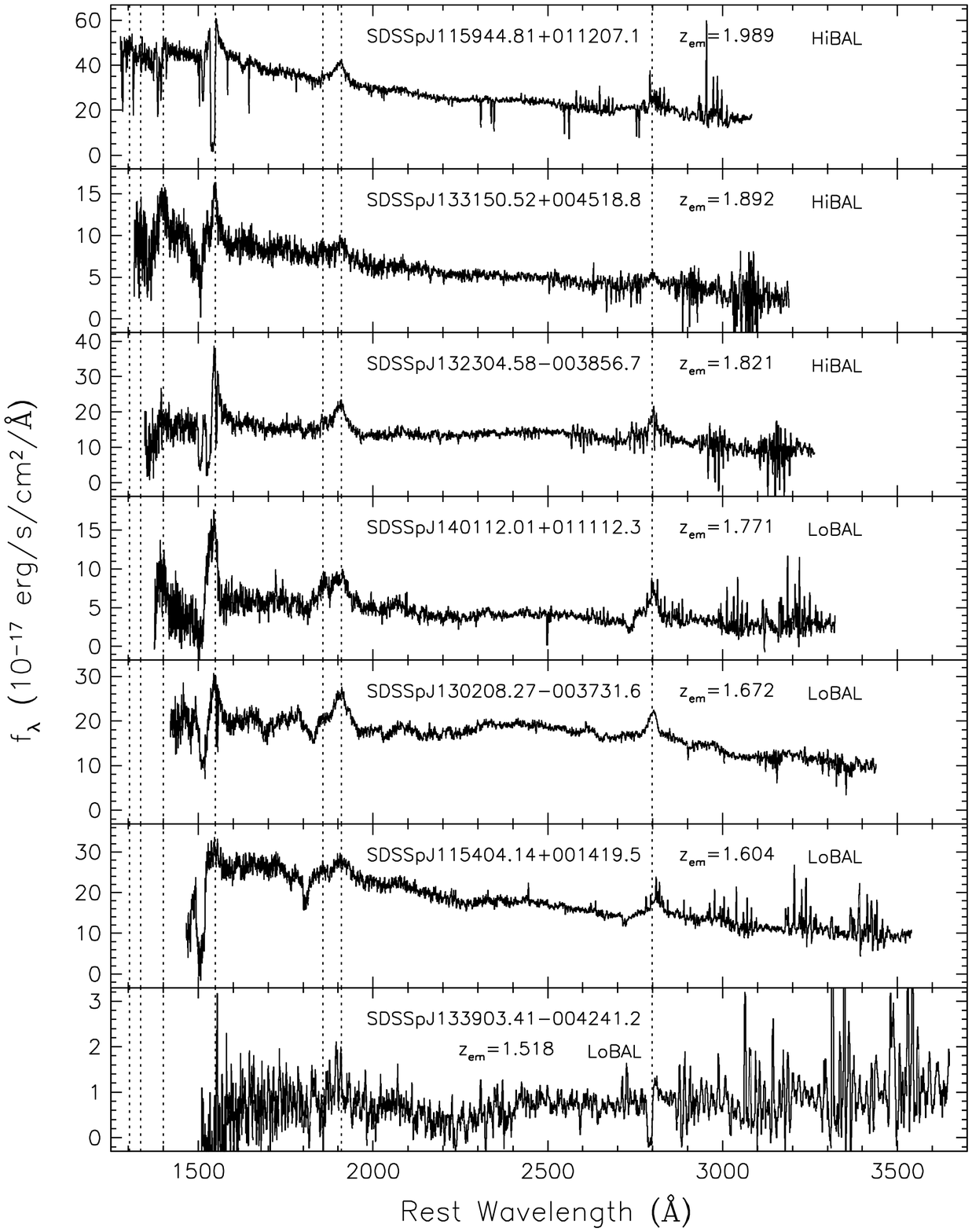}
\caption{{\it continued }
}
\end{figure}

\clearpage
\begin{figure}
\plottwo{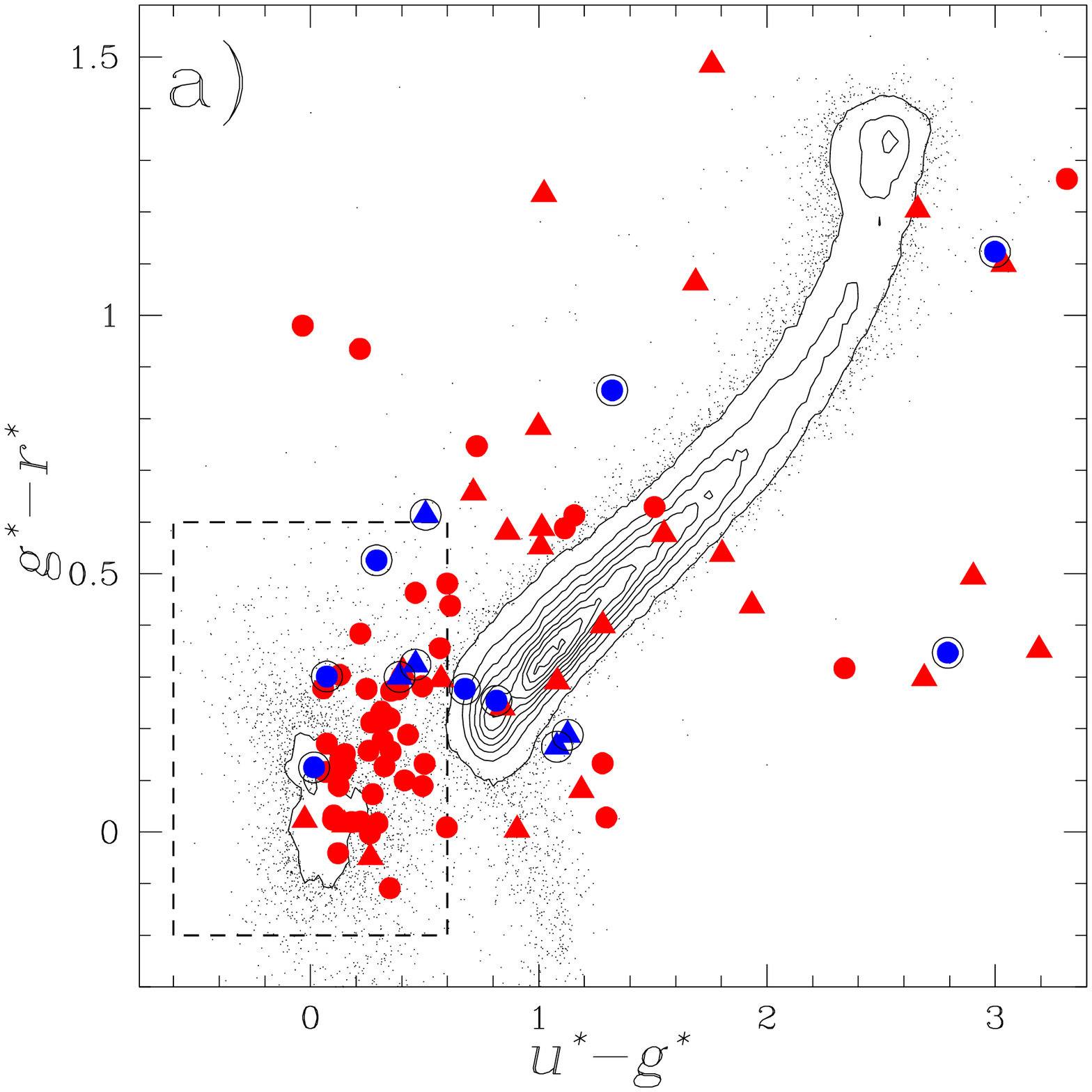}{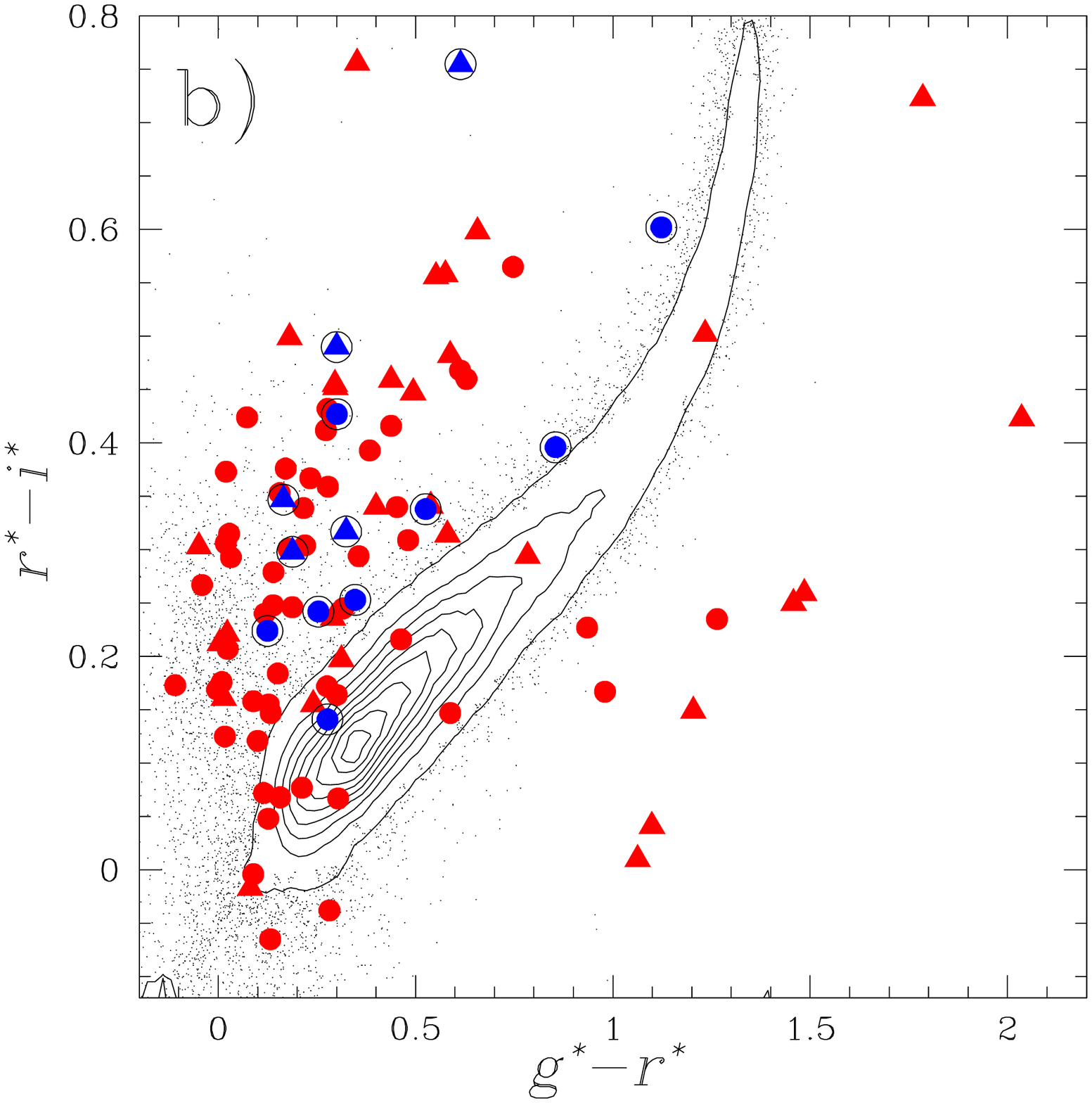}
\caption{(a) Color-color diagram ($u^*-g^*~vs.~g^*-r^*$) for 13
radio-detected BALs (circled symbols) and 96 radio-undetected BALs
(bare symbols). Bright objects ($i^* < 19$) are indicated by filled
circles and faint ones ($i^* \geq 19$) by filled triangles. The
background contours and dots represent the stellar locus for $\sim
40,000$ point-source objects in an SDSS control sample. The dashed
lines show the location of the low-redshift quasar box (see text for
details). (b) The corresponding $g^*-r^*~vs.~r^*-i^*$ color-color
diagram.
\label{fig:two}}
\end{figure}

\clearpage

\begin{figure}
\plottwo{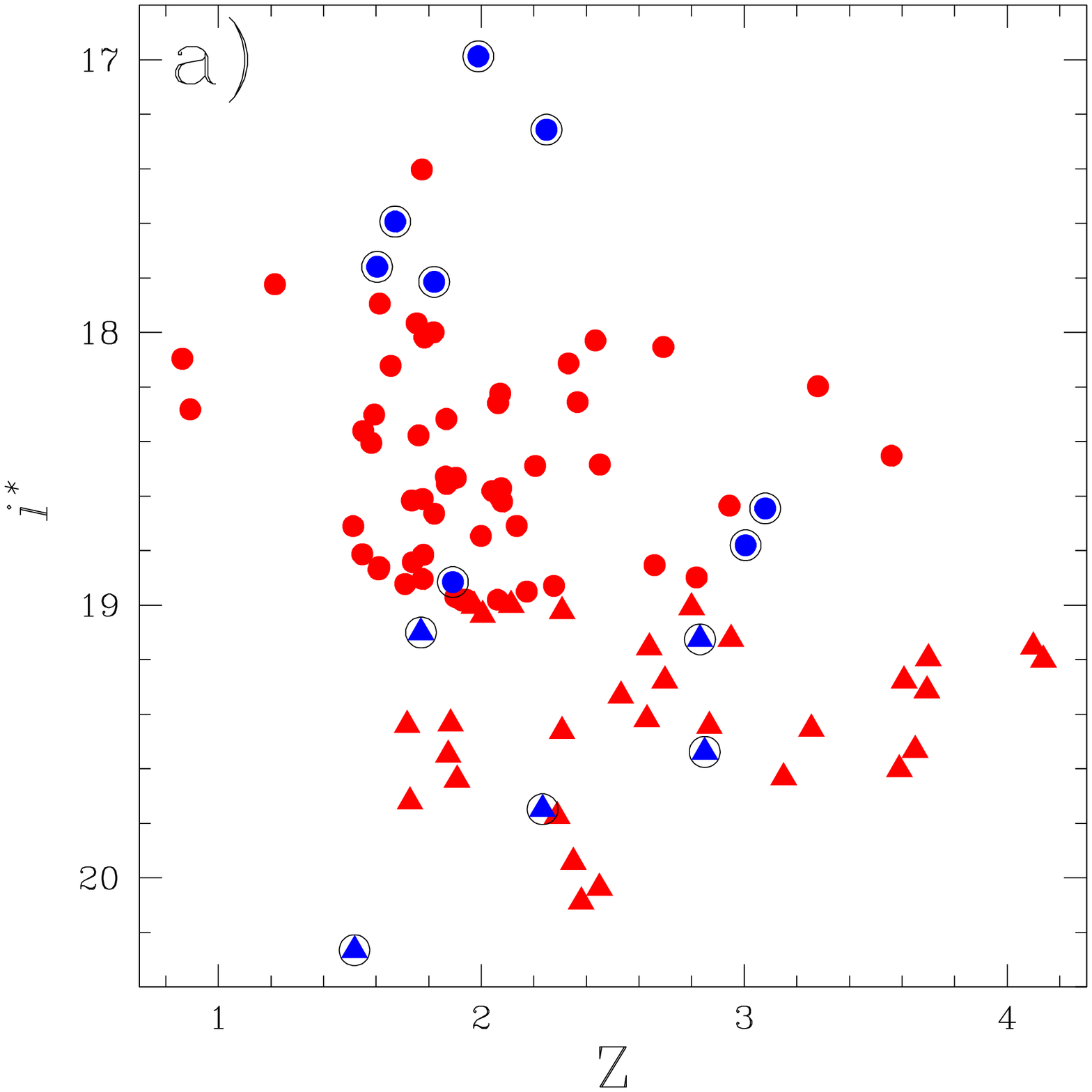}{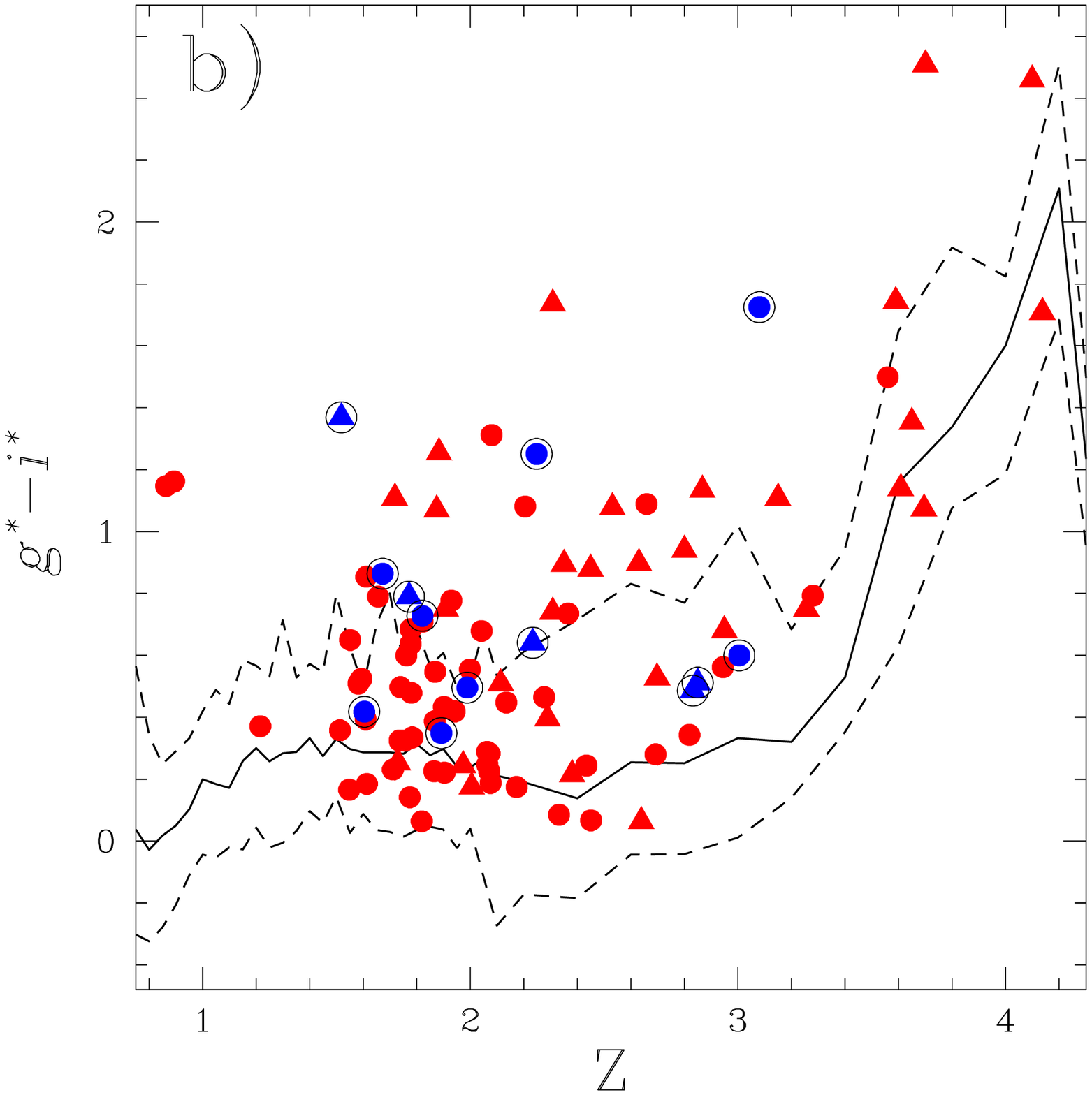}
\caption{Same notation as Fig.~2. (a) Redshift-magnitude ($i^*$)
distribution of the BAL quasars. (b) Color-redshift distribution of
the BAL quasars, compared to the median color-redshift relation (solid
line) derived by Richards et al. (2001a; dashed lines indicate $95\%$
confidence limits). The BAL quasars are redder than the typical
quasars spectroscopically identified by SDSS.
\label{fig:three}}
\end{figure}

\end{document}